\documentclass[preprint,pre,aps,amsfonts,amsmath,superscriptaddress,nofootinbib,showkeys]{revtex4}
\usepackage{amsfonts,amsmath,amsthm,amssymb,amscd}
\usepackage{mathrsfs}
\usepackage{empheq}
\usepackage{graphicx}
\usepackage{bm}
\usepackage{color}
\usepackage[T1]{fontenc}
\newtheorem{condition}{Condition}
\newtheorem{lemma}{Lemma}
\newtheorem{proposition}{Proposition}
\begin{document}
%
%
\title{Finite-Time Transition to Intermittency for a Stochastic Heat Equation Driven by the Square of a Gaussian Field}
\author{Philippe Mounaix}
\email{philippe.mounaix@polytechnique.edu}
\affiliation{CPHT, CNRS, \'Ecole
polytechnique, Institut Polytechnique de Paris, 91120 Palaiseau, France.}
\date{\today}
\begin{abstract}
In this paper, we study the spatial behavior of the solution $\psi(x,t)$ to the stochastic heat equation $\partial_t\psi(x,t)-\frac{1}{2}\partial^2_{x^2} \psi(x,t)=g\, S(x,t)^2\, \psi(x,t)$, with $0\le t\le T$, $x\in\mathbb{R}$, and $\psi(x,0)=1$. Here, $g>0$ is a coupling constant and $S(x,t)$ is a stationary, homogeneous, and ergodic Gaussian field. Focusing on $\mathcal{E}(x,g)\equiv \psi(x,T)$ at a finite time $T>0$, we identify the critical coupling $g_c(T)$ above which the average of $\mathcal{E}(0,g)$ diverges. We show that in the subcritical regime $g<g_c(T)$, $\mathcal{E}(x,g)$ is spatially ergodic, with no intermittency, while in the supercritical regime $g>g_c(T)$ it becomes spatially intermittent and loses ergodicity. Our results differ from the extensively studied case where $S(x,t)^2$ is replaced by $S(x,t)$, in which intermittency appears only asymptotically as $T\to +\infty$, with no finite-time intermittency.
\end{abstract}
\maketitle
%
%
\section{Introduction}\label{sec1}
This paper is devoted to the analysis of the spatial behavior of the solutions to the stochastic heat equation,
\begin{equation}\label{withDeq}
\left\lbrace
\begin{array}{l}
\partial_t\psi(x,t)-\frac{1}{2}\partial^2_{x^2} \psi(x,t)=g\, S(x,t)^2\, \psi(x,t), \\
0\le t\le T,\ x\in\mathbb{R},\ {\rm and}\ \psi(x,0)=1,
\end{array}\right.
\end{equation}
at a fixed finite time $T>0$, depending on the coupling constant $g>0$. In equation~(\ref{withDeq}), $S(x,t)$ denotes a real-valued Gaussian random field whose statistical properties will be specified in Section \ref{sec2a}. For simplicity, we consider $x\in\mathbb{R}$ and we take a constant initial condition $\psi(x,0)=1$. All the results presented here can be straightforwardly generalized to $x\in\mathbb{R}^d$ with $d\ge1$, as well as to homogeneous and ergodic random initial conditions.

The diffraction version of (\ref{withDeq}) -- obtained by replacing $\partial^2_{x^2}$ with $i\partial^2_{x^2}$ -- arose and continues to be of interest in modeling the backscattering of an incoherent high intensity laser light by a plasma. There is a large literature on this topic, and we refer the interested reader to the seminal paper \cite{RD1994}. Considering a diffusive process has the advantage of allowing a relatively straightforward control of the growth of the solution to equation (\ref{withDeq}), something that would be much more challenging to achieve in the diffractive case \cite{ADLM2001,MCL2006}. Besides being interesting in its own right as a mathematical physics problem, the stochastic heat equation (\ref{withDeq}) is thus a natural starting point for approaching the more complex diffraction-amplification problem, whose full solution is still out of reach. To set up the type of questions that will concern us, it is instructive to begin by recalling some known results on sums of independent random variables with heavy-tailed distributions.

Let $U_1,\, U_2,\cdots ,\, U_N$ be a sequence of $N$ independent and identically distributed (i.i.d.) nonnegative random variables with probability distribution function (pdf) $p_U(u)$. Write $\langle U^k\rangle$ ($k\ge 1$) the $k$th (possibly infinite) moment of $p_U(u)$. Assume that $p_U(u)$ decreases like $u^{-(1+\alpha)}$ as $u\to +\infty$, with $\alpha>0$ (heavy-tailed distribution), and consider the sample average
\begin{equation}\label{sampaveriidrv}
\overline{U}_N=\frac{1}{N}\, \sum_{n=1}^N U_n .
\end{equation}
In the large $N$ limit, the behavior of $\overline{U}_N$ falls into three different regimes, depending on the value of $\alpha$. The first two regimes encountered as $\alpha$ is decreased are a classical subject of the theory of probability (see, e.g., \cite{GK1954,Levy1954}):
\begin{description}
\item[regime 1]
\textit{($\, \alpha >2\, $)} In this case, both $\langle U\rangle$ and $\langle U^2\rangle$ exist. As a result, the strong law of large numbers and the central limit theorem apply: $\overline{U}_N$ converges almost surely (a.s.) to $\langle U\rangle$ as $N\to +\infty$, with Gaussian fluctuations that shrink to zero like $N^{-1/2}$.
\item[regime 2]
\textit{($\, 1<\alpha <2\, $)} In this regime, $\langle U\rangle$ is finite while $\langle U^2\rangle$ diverges, so the usual central limit theorem no longer applies and must be replaced by its generalized, stable-law version. The sample mean $\overline{U}_N$ still converges almost surely to $\langle U\rangle$, but its fluctuations are now non-Gaussian and decay like $N^{1/\alpha -1}$ (or $\sqrt{\ln N/N}$ for $\alpha =2$).
\end{description}
In these two regimes -- i.e., for $\alpha>1$ -- the key points for our purposes are that 
\begin{equation}\label{stronglln}
\lim_{N\to +\infty}\overline{U}_N =\langle U\rangle \qquad\textrm{a.s.},
\end{equation}
and that the occurrence of very large values among the $U_n$ is too rare to make a significant contribution to $\overline{U}_N$ as $N\to +\infty$. The sample mean is determined by the overwhelming majority of the $U_n$ with $U_n=O(\langle U\rangle)$.

The third, remaining regime corresponds to a completely different situation:
\begin{description}
\item[regime 3]
\textit{($\, 0<\alpha <1\, $)} In this case, $\langle U\rangle$ is infinite, so the law of large numbers no longer applies. The distribution of the normalized sample mean $\overline{U}_N/N^{1/\alpha -1}$ converges to the stable law $L_{\alpha,1}$ as $N\to +\infty$ \cite{GK1954,Levy1954}. In this limit, $\overline{U}_N$ itself diverges almost surely like $N^{1/\alpha -1}$ (or $\ln N$ when $\alpha =1$) and is asymptotically given by the contribution of the $r_N\sim \ln(\ln N)$ largest values among the $U_n$ \cite{CHM1986,Haeusler1993}. Note that $r_N/N\to 0$ as $N\to +\infty$, which means that these leading $U_n$ constitute a vanishing fraction of the sequence in this limit.
\end{description}
Thus, for $\alpha <1$, the key points are that equation (\ref{stronglln}) is replaced with
\begin{equation}\label{divsampmean}
\lim_{N\to +\infty}\overline{U}_N =+\infty \qquad\textrm{a.s.},
\end{equation}
and that the diverging sample mean is driven by a small minority of the $U_n$ with very large values. Although such extreme values occur only rarely,
\begin{equation}\label{limfractionrv}
\lim_{N\to +\infty}\frac{r_N}{N} =0,
\end{equation}
they provide the dominant contribution to $\overline{U}_N$ as $N\to +\infty$. This behavior defines the \textit{intermittency} of the sequence $U_1,\, U_2,\cdots ,\, U_N$ in the regime $\alpha <1$, that is, the predominance of rare, heavy-tailed extremes in the sample mean $\overline{U}_N$.

The setting for the present work is obtained by replacing the sequence of random variables with a random field $U(x)$ defined over a continuous index $x$ (which may represent a position in space or a moment in time). The assumption that the random variables are identically distributed becomes, in this context, a requirement that the statistics of $U(x)$ remain unchanged under translations in $x$. For $y\ne x$, $U(y)$ and $U(x)$ are generally not independent, so the conditions under which the law of large numbers applies are replaced by those required for the random field $U(x)$ to be ergodic. When these conditions are met, the analogue of (\ref{stronglln}) becomes (assuming for simplicity that $x\in\mathbb{R}$),
\begin{equation}\label{llntoergo}
\lim_{L\to +\infty}\frac{1}{L}\int_{-L/2}^{L/2} U(x)\, dx =\langle U(0)\rangle \qquad\textrm{a.s.},
\end{equation}
as long as the expectation $\langle U(0)\rangle$ exists. Intermittency of $U(x)$ is defined similarly to the random-variable case. In short, $U(x)$ is intermittent if, almost surely, its $x$-average is dominated by rare, very high peaks that occupy only a negligible portion of the domain. More precisely, $U(x)$ is intermittent if there exists a nondecreasing function $f(L)$ with $f(L)\to +\infty$ as $L\to +\infty$ such that, almost surely,
\begin{equation}\label{introfieldinterm1}
\frac{1}{L}\int_{-L/2}^{L/2} U(x)\, dx \sim
\frac{1}{L}\int_{-L/2}^{L/2} U(x)\, \mathbf{1}_{\{ U(x)>f(L)\}}\, dx \qquad (L\to +\infty),
\end{equation}
and
\begin{equation}\label{introfieldinterm2}
\lim_{L\to +\infty}\frac{1}{L}\int_{-L/2}^{L/2} \mathbf{1}_{\{ U(x)>f(L)\}}\, dx =0,
\end{equation}
which is the counterpart of equation (\ref{limfractionrv}).

The connection between heavy-tailed random variables and problem (\ref{withDeq}) relies on the following observation. Since $S$ is Gaussian, its (functional) pdf is the exponential of a negative-definite quadratic form of $S$. In equation (\ref{withDeq}), the amplification of $\psi$ is driven by $S^2$, which in turn brings in an exponential of a positive-definite quadratic form of $S$. Simple scaling arguments then point to an algebraic decay of the pdf $p_{\Psi}(\psi)$ of $\psi(0,T)$, with a tail of the form $\psi^{-(1+\alpha)}$ as $\psi\to +\infty$, with $\alpha >0$. This is consistent with the results of \cite{ADLM2001,MCL2006} on the divergence of $\langle\psi(0,T)^p\rangle$ ($p\in\mathbb{N}$), provided $\alpha =g_c(T)/g$, where the critical coupling $g_c(T)>0$, defined in equation (\ref{critcoupling}), is a decreasing function of $T$. To date, and to the author's knowledge, such an algebraic tail has been established -- through instanton analysis~-- only for the diffraction version of (\ref{withDeq}), obtained by replacing $\partial^2_{x^2}$ with $i\partial^2_{x^2}$  \cite{Mounaix2023,Mounaix2024,Mounaix2025}. Assuming that the same holds in the diffusive case, then $p_{\Psi}(\psi)$ and $p_U(u)$ would share the same heavy upper tail. It is therefore natural to expect that the spatial profile of $\psi(x,T)$ mirrors the fluctuations of the sequence $U_1,\, U_2,\cdots ,\, U_N$ along its index, with the regimes $\alpha >1$ and $\alpha <1$ corresponding respectively to $g<g_c(T)$ and $g>g_c(T)$. In particular, for any finite time $T>0$, one expects $\psi(x,T)$ to be spatially intermittent, in the sense of equations (\ref{introfieldinterm1}) and (\ref{introfieldinterm2}), whenever $g>g_c(T)$. Equivalently, if $g>\min_{T>0} g_c(T)$ is fixed, then there exists a finite critical time $T_c(g)>0$, defined by $T_c(g)=g_c^{-1}(g)$, such that for all $T>T_c(g)$ the field $\psi(x,T)$ should be spatially intermittent. This finite-time transition to intermittency as $g$ exceeds $g_c(T)$ (or as $T$ exceeds $T_c(g)$) parallels the transition to intermittency for sequences of heavy-tailed random variables as $\alpha$ drops below $1$. The goal of this paper is to prove that, in the case of the stochastic heat equation (\ref{withDeq}), this scenario is indeed realized. We will do so without using the upper tail of $p_{\Psi}(\psi)$, whose determination is a separate problem.

Before we begin, it is useful to highlight the following point. The previous discussion of  heavy-tailed random variables shows that the onset of intermittency comes in with diverging moments of $p_U(u)$. In the context of equation (\ref{withDeq}), this translates into diverging moments of $p_{\Psi}(\psi)$. A finite-time transition to intermittency therefore implies that moments blow up at some finite time. This tells us that the finite-time transition considered here is different from the asymptotic intermittency as $T\to +\infty$ \cite{GM1990,Molchanov1991}, which has been extensively studied for the stochastic heat equation (\ref{withDeq}) where $S(x,t)^2$ is replaced by $S(x,t)$; see, e.g., \cite{Mikhailov1989,Mikhailov1991,BC1995,Noble1997,BG1998} and references therein. In that setting, the amplification of $\psi$ involves the exponential of a \textit{linear} form of $S$, leading to a log-normal-type tail of $p_{\Psi}(\psi)$, rather than algebraic. For log-normal-type tails, no moment $\langle\psi(0,T)^p\rangle$ can diverge at any finite time $T>0$ (for any $g>0$, and $p\in\mathbb{N}$). These moments only blow up in the limit $T\to +\infty$. As a result, there is no finite-time intermittency in that case: intermittency only appears asymptotically as $T\to +\infty$. We will not discuss asymptotic intermittency further, except for stressing this point that sets problem (\ref{withDeq}) apart from previously studied models.

The paper is organized as follows. In section \ref{sec2}, we specify the class of $S$ we consider and we determine the corresponding value of the critical coupling $g_c(T)$. Ergodicity and absence of intermittency for $\psi(x,T)$ in the subcritical regime $g<g_c(T)$ are established in section \ref{sec3}. Intermittency and loss of ergodicity in the supercritical regime $g>g_c(T)$ are proved in section \ref{sec4}. Finally, we discuss some open questions and outline directions for future work in section \ref{sec5}. Selected technical material is collected in the appendices.
%
%
\section{Preliminaries}\label{sec2}
\subsection{Specification of $\bm{S(x,t)}$}\label{sec2a}
In Eq.\ (\ref{withDeq}), $S(x,t)$ is a real-valued, homogeneous and stationary Gaussian random field defined by
\begin{equation}\label{defofS}
\begin{array}{l}
\langle S(x,t)\rangle =0, \\
\langle S(x,t)S(x^\prime,t^\prime)\rangle =C(x-x^\prime,t-t^\prime),
\end{array}
\end{equation}
with normalization $C(0,0)=1$. Here and below, homogeneous and stationary mean that $S(\cdot ,\cdot)$ is invariant in law under spatial and temporal shifts, respectively. Stationarity is assumed for simplicity; the case of a nonstationary $S(x,t)$ can be treated similarly, with a little more technical calculations and not substantially different results.

As a zero-mean Gaussian field, $S(x,t)$ is fully characterized by its correlation function $C(x,t)$ or, equivalently, its spectral density $D(k,\omega)$ defined through the Fourier representation
\begin{equation}\label{FourierofC}
C(x,t)=\int\int_{\mathbb{R}^2}D(k,\omega)\, {\rm e}^{i(kx+\omega t)}\, dk\, d\omega.
\end{equation}
Since $C(x,t)$ is the correlation function of a real-valued field, it is positive-definite and real, which implies
\begin{equation}\label{Dproperties}
D(k,\omega)\ge 0,\ \ D(-k,-\omega)=D(k,\omega).
\end{equation}
The class of $S$ considered in this paper is defined by the following two conditions on $D(k,\omega)$:
\begin{condition}\label{condition1}
(Ergodicity). For each $t\in\mathbb{R}$, the spatial spectral density $\int_{\mathbb R} D(k,\omega)\, {\rm e}^{i\omega t}\, d\omega$ is absolutely continuous with respect to Lebesgue measure in $k$. In other words, it has no discrete spectral component, i.e., no term proportional to $\delta(k-k_0)$ for some $k_0\in\mathbb{R}$.
\end{condition}
\begin{condition}\label{condition2}
(Smoothness). $D(k,\omega)$ satisfies
\begin{equation}\label{Dintegrability}
\int\int_{\mathbb{R}^2}(1+\vert k\vert^{2m_1}+\vert\omega\vert^{2m_2})\, D(k,\omega)\, dk\, d\omega <\infty ,
\end{equation}
for some $m_1,\, m_2>2$.
\end{condition}
Condition~\ref{condition1} is equivalent to the Gaussian field $S$ being ergodic with respect to spatial translations \cite{Maruyama1949,Grenander1950,BE1972}. Condition~\ref{condition2} ensures $C(x,t)\in C^4(\mathbb{R}^2)$ and $S\in C^2(\mathbb{R}^2)$ almost surely \cite{Adler1981,Delmas1998}. These conditions are not very restrictive, being fulfilled by most physically relevant fields $S$, in particular in laser-plasma interaction setting.
%
%
\subsection{Expectation value of $\bm{\mathcal{E}(0,g)}$ and critical coupling}\label{sec2b}
For a given realization of $S(x,t)$ and $g>0$, we define $\mathcal{E}(x,g)\equiv \psi(x,T)$, making the dependence on $g$ explicit (and the dependence on $T$ implicit). Using the Feynman--Kac formula for the solution to Eq. (\ref{withDeq}), one has
\begin{equation}\label{FKformula}
\mathcal{E}(x,g)=\int_{B(x,T)}\exp\left[ g\int_0^TS(x(\tau),\tau)^2\, d\tau\right]\, d\mathbb{P}_W[x(\cdot)],
\end{equation}
where $\mathbb{P}_W$ is the Wiener measure on the space of continuous paths in $\mathbb{R}$, $C[0,T]$, and $B(x,T)\subset C[0,T]$ is the set of continuous paths ending at $x$ at time $T$, i.e., $x(T)=x$.

Let $\langle\mathcal{E}(0,g)\rangle_S$ denote the expectation value of $\mathcal{E}(0,g)$, where $\langle\cdot\rangle_S$ stands for the average over the realizations of $S$. We define the critical coupling $g_c(T)\in [0,+\infty]$ by
\begin{equation}\label{critcoupling}
g_c(T)=\inf\{\,  g>0 : \langle\mathcal{E}(0,g)\rangle_S=+\infty\, \} .
\end{equation}
As $T$ is fixed in equation (\ref{withDeq}), we will write $g_c$ for $g_c(T)$ throughout up to section \ref{sec4}. Note that $\mathcal{E}(0,g)$ is nondecreasing in $g$, so $\langle\mathcal{E}(0,g)\rangle_S=+\infty$ for all $g>g_c$. To determine $g_c$ we must therefore identify the values of $g$ for which $\langle\mathcal{E}(0,g)\rangle_S$ is finite and those for which $\langle\mathcal{E}(0,g)\rangle_S =+\infty$. The integrand on the right-hand side of Eq.~(\ref{FKformula}) is positive and by Tonelli's theorem, the Wiener and $S$-averages can be interchanged, yielding
\begin{equation}\label{averageampli}
\langle\mathcal{E}(0,g)\rangle_S=\int_{B(0,T)}\mathcal{A}[x(\cdot); g]\, d\mathbb{P}_W[x(\cdot)],
\end{equation}
where
\begin{equation}\label{pathampli}
\mathcal{A}[x(\cdot); g]=\left\langle\exp\left[ g\int_0^TS(x(\tau),\tau)^2\, d\tau\right]\right\rangle_S.
\end{equation}
One is thus led to examine the finiteness of $\mathcal{A}[x(\cdot); g]$ with respect to both $g$ and $x(\cdot)$.

Along these lines, the determination of the critical coupling leads to the following proposition, which extends the result of \cite{MCL2006} to any real-valued Gaussian field $S(x,t)$ with finite normalization $C(0,0)<+\infty$ and spectral density satisfying Condition \ref{condition1}. The detailed proof, given in appendix \ref{newappA}, builds on the approach of \cite{ADLM2001} and completes the argument.
\begin{proposition}\label{criticalcoupling}
$g_c=1/2\mu_{\rm max}$, with $\mu_{\rm max}=\sup_{x(\cdot)\in B(0,T)}\mu_1[x(\cdot)]$,
\end{proposition}
\noindent where $\mu_1[x(\cdot)]$ is the largest eigenvalue of $\hat{T}_{x(\cdot)}$, the covariance operator associated with the process $S(x(\cdot),\cdot)$ (see appendix \ref{newappA} for details).

The rest of the paper is devoted to studying the ergodicity and intermittency of $\mathcal{E}(x,g)$ depending on whether $g$ is below or above $g_c$.
%
%
\section{Subcritical regime ($\bm{g<g_c}$): ergodicity of $\bm{\mathcal{E}(x,g)}$ and absence of intermittency}\label{sec3}
The main features of the spatial behavior of $\mathcal{E}(x,g)$ in the subcritical regime, $g<g_c$, are summarized in the following proposition:
\begin{proposition}\label{ergodicityofE}
For every $g<g_c$, $\langle\mathcal{E}(0,g)\rangle_S <+\infty$ and
\begin{equation}\label{ergodicequation}
\lim_{L\to +\infty}\frac{1}{L}\int_{-L/2}^{L/2} \mathcal{E}(x,g)\, dx =\langle\mathcal{E}(0,g)\rangle_S\qquad \text{a.s.}.
\end{equation}
Furthermore, for any nondecreasing function $f(L)\underset{L\to +\infty}{\uparrow} +\infty$,
\begin{equation}\label{nointermequation}
\lim_{L\to +\infty}\frac{1}{L}\int_{-L/2}^{L/2} \mathcal{E}(x,g)\, \mathbf{1}_{\{ \mathcal{E}(x,g)>f(L)\}}\, dx =0\qquad \text{a.s.}.
\end{equation}
\end{proposition}

\bigskip
Proposition \ref{ergodicityofE} expresses that, in the subcritical regime, the high peaks of $\mathcal{E}(x,g)$ are too rare to contribute to $\langle\mathcal{E}(x,g)\rangle_S$, which therefore remains finite. As $L\to +\infty$, and for almost all realizations of $S$, these peaks are also too scarce to contribute to the space average of $\mathcal{E}(x,g)$, which thus coincides with the ensemble average $\langle\mathcal{E}(0,g)\rangle_S$ (equation (\ref{ergodicequation})). The absence of any contribution from the high peaks to the space average means that $\mathcal{E}(x,g)$ is \textit{not} intermittent in the subcritical regime (equation (\ref{nointermequation}), to be compared with equation (\ref{introfieldinterm1})). More precisely, for any given $0<\gamma <g_c$ (arbitrarily close to $g_c$), proposition \ref{ergodicityofE} shows that $\mathcal{E}(x,g)$ is ergodic and non-intermittent over the entire range $g\le\gamma$. Whether this result can be extended to $g=g_c$ depends on whether $\langle \mathcal{E}(0,g_c)\rangle_S <+\infty$. We will return to this point in Sec. \ref{sec5}. We now give the proof of proposition \ref{ergodicityofE}.

\bigskip
\noindent\textbf{Proof.}
\textit{(First part.)} For every $g<g_c$, $\langle \mathcal{E}(0,g)\rangle_S$ is finite by the definition of $g_c$. To prove the ergodic theorem (\ref{ergodicequation}) for $\mathcal{E}(x,g)$, it is convenient to make the dependence on $S$ explicit (and the dependence on $g$ implicit) by writing $\mathcal{E}(x,g)\equiv\mathcal{E}(x;S)$. The field $\mathcal{E}(x;S)$ is obtained from $S$ through a measurable operation that commutes with spatial translations. Indeed, if $t_a$ denotes the spatial shift by $a$, one checks directly from the Feynman--Kac formula (\ref{FKformula}) that
\begin{equation}\label{equivariantE}
\mathcal{E}(x+a;S)=\mathcal{E}(x;t_a S) ,
\end{equation}
that is, shifting the observation point is equivalent to shifting the underlying field $S$ by the same amount. In the language of ergodic theory, $\mathcal{E}$ is a factor of the translation action on $S$, and standard results ensure that ergodicity is preserved under such factor maps; see, e.g., \cite{Walters1975,Walters1982,CFSS1982,Petersen1989}. Since $S$ is ergodic (by condition \ref{condition1}), it follows that $\mathcal{E}$ is also ergodic for every $g<g_c$, which proves the first part of the proposition.

\bigskip
\textit{(Second part.)} Fix $u>0$ and define $\mathcal{E}_u(x,g)=\mathcal{E}(x,g)\, \mathbf{1}_{\{ \mathcal{E}(x,g)\le u\}}$. From the obvious inequality $\mathcal{E}_u(x,g)\le u$ and the first part of the proof applied to $\mathcal{E}_u$ instead of $\mathcal{E}$, it follows respectively that $\langle\mathcal{E}_u(0,g)\rangle_S\le u <+\infty$ and $\mathcal{E}_u$ is homogeneous and ergodic, hence
\begin{equation}\label{ergotruncE}
\lim_{L\to +\infty}\frac{1}{L}\int_{-L/2}^{L/2} \mathcal{E}_u(x,g)\, dx =\langle\mathcal{E}_u(0,g)\rangle_S\qquad \text{a.s.}.
\end{equation}
Since $f(L)\underset{L\to +\infty}{\uparrow} +\infty$, for all $L$ large enough one has $f(L)>u$ and
\begin{equation*}
\frac{1}{L}\int_{-L/2}^{L/2} \mathcal{E}_{f(L)}(x,g)\, dx \ge \frac{1}{L}\int_{-L/2}^{L/2} \mathcal{E}_u(x,g)\, dx ,
\end{equation*}
which yields, using equation (\ref{ergotruncE}),
\begin{equation}\label{EfLinf1}
\liminf_{L\to +\infty}\frac{1}{L}\int_{-L/2}^{L/2} \mathcal{E}_{f(L)}(x,g)\, dx \ge\langle\mathcal{E}_u(0,g)\rangle_S\qquad \text{a.s.}.
\end{equation}
For finite $\mathcal{E}(0,g)$ and for $u$ large enough, one has $\mathcal{E}(0,g)\le u$, and $\lim_{u\to +\infty}\mathcal{E}_u(0,g)=\mathcal{E}(0,g)$. On the other hand, if $\mathcal{E}(0,g)= +\infty$, then $\mathcal{E}(0,g)> u$ for all $u$, and $\lim_{u\to +\infty}\mathcal{E}_u(0,g)=0$. Since $\langle\mathcal{E}(0,g)\rangle_S <+\infty$, one has $\textrm{Prob.}(\mathcal{E}(0,g)= +\infty)=0$ and
\begin{equation*}
\mathcal{E}_u(0,g)\underset{u\to +\infty}{\uparrow} \mathcal{E}(0,g)\qquad\textrm{a.s.}.
\end{equation*}
As a result, the monotone convergence theorem yields
\begin{equation*}
\langle\mathcal{E}_u(0,g)\rangle_S\underset{u\to +\infty}{\uparrow} \langle\mathcal{E}(0,g)\rangle_S,
\end{equation*}
and letting $u$ arbitrarily large on the right-hand side of (\ref{EfLinf1}) gives
\begin{equation}\label{EfLinf2}
\liminf_{L\to +\infty}\frac{1}{L}\int_{-L/2}^{L/2} \mathcal{E}_{f(L)}(x,g)\, dx \ge\langle\mathcal{E}(0,g)\rangle_S\qquad \text{a.s.}.
\end{equation}
An upper bound is readily obtained from $\mathcal{E}_{f(L)}(x,g)\le \mathcal{E}(x,g)$ and the limit (\ref{ergodicequation}), yielding
\begin{equation}\label{EfLsup}
\limsup_{L\to +\infty}\frac{1}{L}\int_{-L/2}^{L/2} \mathcal{E}_{f(L)}(x,g)\, dx \le\langle\mathcal{E}(0,g)\rangle_S\qquad \text{a.s.}.
\end{equation}
Combining inequalities (\ref{EfLinf2}) and (\ref{EfLsup}), one obtains
\begin{equation}\label{complnointerm}
\lim_{L\to +\infty}\frac{1}{L}\int_{-L/2}^{L/2} \mathcal{E}_{f(L)}(x,g)\, dx =\langle\mathcal{E}(0,g)\rangle_S\qquad \text{a.s.}.
\end{equation}
Equation (\ref{complnointerm}), together with (\ref{ergodicequation}), is equivalent to the limit (\ref{nointermequation}), which completes the proof of the proposition.
%
%
\section{Supercritical regime ($\bm{g>g_c}$): intermittency of $\bm{\mathcal{E}(x,g)}$ and loss of ergodicity}\label{sec4}
The specific properties of the spatial behavior of $\mathcal{E}(x,g)$ in the supercritical regime, $g>g_c$, are summarized in the following proposition:
\begin{proposition}\label{intermittencyofE}
For every $g>g_c$,
\begin{equation}\label{divspaceaverage}
\lim_{L\to +\infty}\frac{1}{L}\int_{-L/2}^{L/2} \mathcal{E}(x,g)\, dx =+\infty \qquad \text{a.s.},
\end{equation}
and
\begin{equation}\label{noergodicityquation}
\lim_{L\to +\infty}\frac{1}{L\langle\mathcal{E}(0,g)\rangle_S}\int_{-L/2}^{L/2} \mathcal{E}(x,g)\, dx =0\qquad \text{a.s.}.
\end{equation}
Furthermore, there exists a nondecreasing function $f(L)\underset{L\to +\infty}{\uparrow} +\infty$ such that, almost surely,
\begin{equation}\label{intermittentasym}
\frac{1}{L}\int_{-L/2}^{L/2} \mathcal{E}(x,g)\, \mathbf{1}_{\{ \mathcal{E}(x,g)>f(L)\}}\, dx
\sim \frac{1}{L}\int_{-L/2}^{L/2} \mathcal{E}(x,g)\, dx \qquad (L\to +\infty),
\end{equation}
and
\begin{equation}\label{domdomainsize}
\lim_{L\to +\infty}\frac{1}{L}\int_{-L/2}^{L/2} \mathbf{1}_{\{ \mathcal{E}(x,g)>f(L)\}}\, dx =0.
\end{equation}
\end{proposition}

\bigskip
The meaning of proposition \ref{intermittencyofE} is that, in the supercritical regime $g>g_c$, $\mathcal{E}(x,g)$ is intermittent, in the sense of equations (\ref{introfieldinterm1}) and (\ref{introfieldinterm2}). Namely, as $L\to +\infty$ and for almost all realizations of $S$, the space average of $\mathcal{E}(x,g)$ is dominated entirely by the contribution of the highest peaks (equation (\ref{intermittentasym})), even though these peaks occupy only a negligible portion of space (equation (\ref{domdomainsize})). In this regime, the scarcity of the high peaks of $\mathcal{E}(x,g)$ is outweighed by their extremely large amplitudes. At the same time, for any arbitrarily large $L>0$, the space average of a single realization fails to sample the values of $\mathcal{E}(x,g)$ exceeding the maximum over $[-L/2,\, L/2]$ -- that are responsible for the divergence of $\langle\mathcal{E}(0,g)\rangle_S$ -- leading to the breakdown of ergodicity (equation (\ref{noergodicityquation})). The proof of proposition \ref{intermittencyofE} is as follows.

\bigskip
\noindent\textbf{Proof.}
\textit{(First part.)}
Fix $u>0$ and consider again $\mathcal{E}_u(x,g)=\mathcal{E}(x,g)\, \mathbf{1}_{\{ \mathcal{E}(x,g)\le u\}}$. For every $u>0$ and every $L>0$, the counterpart of equation (\ref{EfLinf1}) for $\mathcal{E}$ instead of $\mathcal{E}_{f(L)}$, reads
\begin{equation}\label{limintEgtintEu}
\liminf_{L\to +\infty}\frac{1}{L}\int_{-L/2}^{L/2} \mathcal{E}(x,g)\, dx \ge\langle\mathcal{E}_u(0,g)\rangle_S\qquad \text{a.s.}.
\end{equation}
Now, to proceed, we need to modify the argument below equation (\ref{EfLinf1}) so as not to use $\langle\mathcal{E}(0,g)\rangle_S<+\infty$, which fails for all $g>g_c$.

By Condition~\ref{condition2}, $S(x,t)$ is almost surely $C^2$ in $\mathbb{R}^2$, and parabolic regularity \cite{Winiarska1992} then implies that $\psi(x,t)$ is almost surely $C^2$ in $\mathbb{R}\times (0,T]$. In particular, $\psi(x,t)$ is almost surely bounded on any compact subset of $\mathbb{R}\times (0,T]$, so that $\mathcal{E}(x,g)\equiv \psi(x,T)$ is almost surely finite for every given $x\in\mathbb{R}$. Consequently, $\textrm{Prob.}(\mathcal{E}(0,g)= +\infty)=0$ and the same argument used below equation (\ref{EfLinf1}) shows that
\begin{equation*}
\mathcal{E}_u(0,g)\underset{u\to +\infty}{\uparrow} \mathcal{E}(0,g)\qquad\textrm{a.s.},
\end{equation*}
in the supercritical regime as well. Finally, since $\langle\mathcal{E}(0,g)\rangle_S=+\infty$ for all $g>g_c$, the monotone convergence theorem implies
\begin{equation*}
\langle\mathcal{E}_u(0,g)\rangle_S\underset{u\to +\infty}{\uparrow} +\infty ,
\end{equation*}
and letting $u\to +\infty$ in (\ref{limintEgtintEu}) yields
\begin{equation}\label{divinfspaceaverage}
\liminf_{L\to +\infty}\frac{1}{L}\int_{-L/2}^{L/2} \mathcal{E}(x,g)\, dx =+\infty \qquad \text{a.s.},
\end{equation}
hence the limit (\ref{divspaceaverage}).

The fact that $\psi(x,t)$ is almost surely bounded on any compact subset of $\mathbb{R}\times (0,T]$ ensures that, for any $L>0$, $\sup_{x\in[-L/2,L/2]}\mathcal{E}(x,g)<+\infty$ almost surely, and therefore
\begin{equation}\label{asboundedintegral1}
I_L=\int_{-L/2}^{L/2} \mathcal{E}(x,g)\, dx \le L\sup_{x\in[-L/2,L/2]}\mathcal{E}(x,g)<+\infty \qquad\textrm{a.s.}.
\end{equation}
In particular, equation (\ref{asboundedintegral1}) yields $\textrm{Prob.}(I_n <+\infty)=1$ for every $n\in\mathbb{N}$. Define the events
\begin{equation}\label{twoeventsdef}
\begin{array}{l}
A=\{\forall n\in\mathbb{N},\ I_n<+\infty\}=\cap_{n=1}^{+\infty}\{I_n<+\infty\}, \\
B=\{\forall L>0\ (L\in\mathbb{R}),\ I_L<+\infty\}.
\end{array}
\end{equation}
Since $A$ is a countable intersection of full-measure events, one has $\textrm{Prob.}(A)=1$. Fix a realization in $A$. For any $L>0$, choose $n\in\mathbb{N}$ with $n>L$. As $\mathcal{E}(x,g)\ge 0$, we have $I_L\le I_n <+\infty$ for this realization. Thus, every realization in $A$ also lies in $B$, so $A\subseteq B$ and therefore $\textrm{Prob.}(B)\ge\textrm{Prob.}(A)$. (In fact, since (\ref{twoeventsdef}) implies $B\subseteq A$, one has equality.) Hence
\begin{equation}\label{asboundedintegral2}
\textrm{Prob.}(\forall L>0,\ I_L<+\infty)=1.
\end{equation}
Finally, as $\langle\mathcal{E}(0,g)\rangle_S=+\infty$ for all $g>g_c$, equation (\ref{asboundedintegral2}) yields
\begin{equation}\label{asergodicityloss1}
\textrm{Prob.}\left(\forall L>0,\ \frac{I_L}{L\langle\mathcal{E}(0,g)\rangle_S}=0\right)=1,
\end{equation}
and since the event in (\ref{asergodicityloss1}) is contained in the event $\{\lim_{L\to +\infty}I_L/(L\langle\mathcal{E}(0,g)\rangle_S)=0\}$, one obtains
\begin{equation}\label{asergodicityloss2}
\textrm{Prob.}\left(\lim_{L\to +\infty}\frac{I_L}{L\langle\mathcal{E}(0,g)\rangle_S}=0\right)=1,
\end{equation}
which coincides with the limit (\ref{noergodicityquation}).

\bigskip
\textit{(Second part.)}
Let $\phi(u)$ be a nondecreasing function such that  $\phi(u)\to +\infty$ as $L\to +\infty$, and $\lim_{u\to +\infty}\phi(u)/u=0$. Define $f(L)=\phi(\inf_{\ell\ge L}(I_\ell/\ell))$. Clearly, from equation (\ref{divinfspaceaverage}) and the properties of $\phi(u)$, it follows that $f(L)$ is a nondecreasing function with $f(L)\to +\infty$ as $L\to +\infty$, and
\begin{equation}\label{fisoofIL}
\lim_{L\to +\infty}\left(\frac{1}{L\, f(L)}\int_{-L/2}^{L/2}\mathcal{E}(x,g)\, dx\right)^{-1} =0 \qquad\textrm{a.s.}.
\end{equation}
One has the inequalities
\begin{equation}\label{chainofineq1}
\frac{1}{L}\int_{-L/2}^{L/2}\mathcal{E}(x,g)\mathbf{1}_{\{\mathcal{E}(x,g)\le f(L)\}}\, dx \le
\frac{f(L)}{L}\int_{-L/2}^{L/2}\mathbf{1}_{\{\mathcal{E}(x,g)\le f(L)\}}\, dx\le f(L).
\end{equation}
Dividing both sides of (\ref{chainofineq1}) by the space average of $\mathcal{E}(x,g)$ over the interval $[-L/2,L/2]$, and taking the limit $L\to +\infty$, using (\ref{fisoofIL}), one obtains
\begin{equation}\label{compintermasym}
\lim_{L\to +\infty}
\left(\frac{1}{L}\int_{-L/2}^{L/2}\mathcal{E}(x,g)\mathbf{1}_{\{\mathcal{E}(x,g)\le f(L)\}}\, dx\right)
\left(\frac{1}{L}\int_{-L/2}^{L/2}\mathcal{E}(x,g)\, dx\right)^{-1} =0 \qquad\textrm{a.s.},
\end{equation}
which proves the asymptotics in equation (\ref{intermittentasym}).

We now proceed to prove equation (\ref{domdomainsize}). Fix $\varepsilon>0$ and consider the following estimate:
\begin{equation}\label{obviousestim}
\frac{1}{L}\int_{-L/2}^{L/2}\mathbf{1}_{\{\mathcal{E}(x,g)>f(L)\}}\, dx \le
\frac{1}{L\, f(L)^{\varepsilon}}\int_{-L/2}^{L/2}\mathcal{E}(x,g)^{\varepsilon} \, dx.
\end{equation}
Assume that for $\varepsilon >0$ small enough, $\langle\mathcal{E}(0,g)^{\varepsilon}\rangle_S <+\infty$ (this assumption will be verified shortly). In this case, by the first part of the proof of proposition \ref{ergodicityofE} applied to $\mathcal{E}^{\varepsilon}$ instead of $\mathcal{E}$, it follows that $\mathcal{E}^{\varepsilon}$ is homogeneous and ergodic. Therefore,
\begin{equation}\label{ergoEepsilon}
\lim_{L\to +\infty}\frac{1}{L}\int_{-L/2}^{L/2} \mathcal{E}(x,g)^{\varepsilon}\, dx =\langle\mathcal{E}(0,g)^{\varepsilon}\rangle_S <+\infty \qquad \text{a.s.},
\end{equation}
and since $f(L)\to+\infty$ as $L\to +\infty$, taking the large $L$ limit of (\ref{obviousestim}) yields
\begin{equation}\label{neglidomain}
\limsup_{L\to +\infty}\frac{1}{L}\int_{-L/2}^{L/2}\mathbf{1}_{\{\mathcal{E}(x,g)>f(L)\}}\, dx \le
\frac{\langle\mathcal{E}(0,g)^{\varepsilon}\rangle_S}{\lim_{L\to +\infty}f(L)^{\varepsilon}} =0 \qquad\textrm{a.s.},
\end{equation}
which proves equation (\ref{domdomainsize}).

\bigskip
It remains to verify that $\langle\mathcal{E}(0,g)^{\varepsilon}\rangle_S <+\infty$ for $\varepsilon$ small enough. As long as the suspected algebraic tail of the pdf of $\mathcal{E}(0,g)$ has not been proved, it cannot be used to establish this result and we need a different approach. To this end, we will use the following lemma that gives the necessary control of $\mathcal{E}(0,g)$:
\begin{lemma}\label{neededbound}
For every realization of $S$, there exists a random variable $M\equiv M(S)>0$ such that
\begin{equation}\label{neededboundeq}
\mathcal{E}(0,g) \le\exp(gTM),
\end{equation}
and whose probability distribution is absolutely continuous, with upper tail pdf
\begin{equation}\label{uppertailpdf}
p_M(m)\sim A\, m^{1/2}\, \exp\!\left(-\frac{m}{2}\right)\qquad (m\to +\infty),
\end{equation}
where $A>0$ is a constant.
\end{lemma}

\bigskip
To avoid interrupting the proof of proposition \ref{intermittencyofE} with a technical detour, we defer the proof of this lemma to Appendix \ref{newappB} and continue with its application to conclude the argument. By equations~(\ref{neededboundeq}) and (\ref{uppertailpdf}),
\begin{equation}\label{finalbound}
\langle\mathcal{E}(0,g)^{\varepsilon}\rangle_S \le \int_0^{+\infty}\, p_M(m)\, \textrm{e}^{\varepsilon  gTm}\, dm,
\end{equation}
where the integral is finite whenever $\varepsilon <(2gT)^{-1}$. Hence $\langle\mathcal{E}(0,g)^{\varepsilon}\rangle_S <+\infty$ for all such $\varepsilon$, which completes the proof of proposition \ref{intermittencyofE}.
%
%
\section{Discussion and perspectives}\label{sec5}
In this paper, we have examined the spatial behavior of the solution $\psi(x,t)$ to a stochastic heat equation driven by the square of a homogeneous and ergodic Gaussian field. Considering $\mathcal{E}(x,g)\equiv \psi(x,T)$ at a fixed time $T>0$ as a function of the coupling constant $g>0$, we have first determined the critical value $g_c(T)$ above which $\langle\mathcal{E}(0,g)\rangle_S$ diverges, extending  previous results \cite{MCL2006} to a wider class of Gaussian driving fields (proposition \ref{criticalcoupling}). We have then shown that in the subcritical regime $g<g_c(T)$, $\mathcal{E}(x,g)$ is spatially ergodic, with absence of intermittency (proposition \ref{ergodicityofE}). Finally, in the complementary supercritical domain $g>g_c(T)$, we have established that $\mathcal{E}(x,g)$ is spatially intermittent, with loss of ergodicity (proposition~\ref{intermittencyofE}).

All the material in sections \ref{sec2} to \ref{sec4} can be straightforwardly extended, \textit{mutatis mutandis}, to diffusion in $\mathbb{R}^d$ with $d\ge 1$. In particular, the conclusions drawn in sections \ref{sec3} and \ref{sec4} for $d=1$ remain valid in higher spatial dimensions: for any $d\ge 1$, there is a finite-time transition from an ergodic, non-intermittent $\mathcal{E}(x,g)$ to an intermittent, non-ergodic $\mathcal{E}(x,g)$ as $g$ exceeds the critical coupling $g_c(T)$ at which $\langle\mathcal{E}(0,g)\rangle_S$ diverges.

The ergodic (non-intermittent) and intermittent (non-ergodic) regimes established in propositions \ref{ergodicityofE} and \ref{intermittencyofE} are well defined only in the limit $L\to\infty$. From the practical viewpoint, an important question is how rapidly these asymptotic regimes are approached as $L$ increases to large (but finite) values. To the best of the author's knowledge, this issue can only be addressed through numerical investigations, which are themselves technically challenging. In the intermittent regime, obtaining a good sampling of the rare, highest peaks that dominate the space average of $\mathcal{E}(x,g)$ is numerically demanding. First, Equation (\ref{withDeq}) must be solved numerically for each realization of $S$ and for a very large system size $L$. This computation must then be repeated for a very large number of realizations of $S$ in order to obtain reliable statistics of $\mathcal{E}(x,g)$. Finally, the entire procedure must be repeated for several increasing values of $L$. A brute-force implementation of this program would require a prohibitively large number of numerical solutions of Equations (\ref{withDeq}). A possible way to circumvent this difficulty would be to bias the distribution of $S$ towards the realizations that dominate the space average of $\mathcal{E}(x,g)$, for example by developing an importance sampling algorithm specifically tailored to the present problem. Even if such an approach proved feasible -- which is far from obvious a priori -- it would still require massive numerical simulations. Undertaking such a numerical study would be of considerable interest, but it lies beyond the scope of the present work.

Returning to the asymptotic limit $L\to\infty$, how intermittency sets in as $g$ increases depends on how $\langle\mathcal{E}(0,g)\rangle_S$ behaves as $g$ approaches $g_c(T)$ from below. If $\langle\mathcal{E}(0,g)\rangle_S$ diverges as $g\uparrow g_c(T)$, then the profile of $\mathcal{E}(x,g)$ increasingly resembles that of an intermittent field as $g$ approaches $g_c(T)$. As stated in proposition \ref{ergodicityofE}, $\mathcal{E}(x,g)$ is \textit{not} intermittent in this regime. This growing resemblance within the subcritical, ergodic regime is a subcritical precursor, revealing a gradual transition to supercritical intermittency which is completed at $g=g_c(T)$. In this case, the supercritical regime extends to all $g\ge g_c(T)$. This scenario is very similar to the one of asymptotic intermittency as $T\to +\infty$ mentioned at the end of section \ref{sec1} \cite{GM1990,Molchanov1991,Mikhailov1989,Mikhailov1991,BC1995,Noble1997,BG1998}, with the limit $T\to +\infty$ replaced here by $g\to g_c(T)$ from below at fixed $T>0$, or equivalently by $T\to T_c(g)$ from below at fixed $g>\min_{T>0}g_c(T)$. If, on the other hand, $\langle\mathcal{E}(0,g)\rangle_S$ remains finite at $g=g_c(T)$, then there is no precursor of intermittency in the subcritical regime, which in this case extends to all $g\le g_c(T)$. Intermittency then appears abruptly as $g$ exceeds $g_c(T)$, resulting in a discontinuous transition. Deciding which scenario is actually realized requires determining whether $\langle\mathcal{E}(0,g_c(T))\rangle_S$ is finite. Technically, this amounts to checking whether the singularity of $\mathcal{A}[x(\cdot); g_c(T)]$ at $\mu_1[x(\cdot)]=\mu_{\rm max}$, which appears in expression (\ref{pathampliestim}) with $g=g_c(T)=1/2\mu_{\rm max}$, is Wiener-integrable. This is a nontrivial open problem that we leave for future investigation.

Still regarding problem (\ref{withDeq}), let us also mention, as an interesting open question, the determination of the optimal $f(L)$ in the second part of proposition \ref{intermittencyofE}, in the same spirit as the optimal $r_N$ in the heavy-tailed random variable problem \cite{CHM1986,Haeusler1993} (see section \ref{sec1}). That is, the largest $f(L)$ satisfying equations (\ref{intermittentasym}) and (\ref{domdomainsize}).

In the long run, we would like to extend the results obtained in this paper to the far more challenging diffractive version of problem (\ref{withDeq}), obtained by replacing $\partial^2_{x^2}$ with $i\partial^2_{x^2}$. The diffractive case is considerably harder because the analogue of (\ref{FKformula}) is no longer well defined as a measure over continuous paths. More precisely, the main difficulty lies in controlling the complex Feynman path integral that replaces the well-defined Wiener integral in the Feynman-Kac formula (\ref{FKformula}). In this regard, it is far from obvious whether the distributional formulation of the Feynman path integral developed in \cite{MCL2006} to determine the critical coupling remains suitable for studying the transition to intermittency. Overcoming this difficulty would be a major step toward understanding intermittency in the diffractive setting relevant, for example, to laser-matter interaction physics.
%
%
\appendix
\section{Proof of proposition \ref{criticalcoupling}}\label{newappA}
We begin with a technical lemma that will be useful in the following. For any given $x(\cdot)\in B(0,T)$, let $\hat{T}_{x(\cdot)}$ denote the covariance operator of the process $S(x(\cdot),\cdot)$, defined by
\begin{equation}\label{covariancepath}
(\hat{T}_{x(\cdot)} f)(t) =\int_0^T
C(x(t)-x(t^\prime) ,t-t^\prime)\, f(t^\prime)\, dt^\prime ,
\ \ \ f\in L^2([0,T];\mathbb{R}),
\end{equation}
with $L^2([0,T];\mathbb{R})$ the (Hilbert) space of square-integrable real-valued functions on $[0,T]$. Since $\vert C(x,t)\vert\le C(0,0)=1$, $\hat{T}_{x(\cdot)}$ is a compact, self-adjoint operator in $L^2([0,T];\mathbb{R})$. Let $\mu_1[x(\cdot)]\ge\mu_2[x(\cdot)]\ge\cdots\ge 0$ denote the eigenvalues of $\hat{T}_{x(\cdot)}$. Then:
\begin{lemma}\label{mucontinuity}
The mapping $x(\cdot)\mapsto\mu_1[x(\cdot)]$ is continuous on $B(0,T)$ endowed with the uniform norm $\| x(\cdot)\|_\infty =\sup_{0\le t\le T}\vert x(t)\vert$.
\end{lemma}

\bigskip
\noindent\textbf{Proof.}
We first show that $C(x,t)$ is continuous in $x$, uniformly in $t\in[-T,T]$. For every $x\in\mathbb{R}$, $x^\prime\in\mathbb{R}$, and $t\in[-T,T]$,  it follows from the Fourier representation (\ref{FourierofC}) that
\begin{equation}\label{uniformconvC1}
\vert C(x,t)-C(x^\prime ,t)\vert\le
\int\int_{\mathbb{R}^2}2D(k,\omega)\, \left\vert\sin\left[\frac{k(x-x^\prime)}{2}\right]\right\vert\, dk\, d\omega ,
\end{equation}
independent of $t$. The integrand on the right-hand side of Eq.~(\ref{uniformconvC1}) is bounded by $2D(k,\omega)$, which is an integrable function in $L^1(\mathbb{R}^2)$ by Condition \ref{condition1} and the normalization $C(0,0)=1$. Hence, by dominated convergence,
\begin{equation}\label{uniformconvC2}
\lim_{x^\prime \to x}\sup_{\vert t\vert\le T}\vert C(x,t)-C(x^\prime ,t)\vert =0.
\end{equation}
Thus, for every $\varepsilon>0$, there exists $\delta>0$ such that, for any $x(\cdot)$, $x^\prime(\cdot)\in B(0,T)$ satisfying $\sup_{0\le\tau\le T}\vert x(\tau)-x^\prime(\tau)\vert\le\delta$, and for all $t$, $t^\prime\in [0,T]$,
\begin{equation*}
-\frac{\varepsilon}{T}\le
C(x^\prime(t)-x^\prime(t^\prime) , t-t^\prime)-C(x(t)-x(t^\prime) , t-t^\prime)
\le\frac{\varepsilon}{T} .
\end{equation*}
Multiplying by $f(t)f(t^\prime)$, where $f\in L^2([0,T];\mathbb{R})$ satisfies $\| f\|_{L^2([0,T];\mathbb{R})}=1$, gives
\begin{equation*}
-\frac{\varepsilon}{T}\vert f(t)f(t^\prime)\vert\le
f(t)\, C(x^\prime(t)-x^\prime(t^\prime) , t-t^\prime)\, f(t^\prime)- f(t)\, C(x(t)-x(t^\prime) , t-t^\prime)\, f(t^\prime)
\le\frac{\varepsilon}{T}\vert f(t)f(t^\prime)\vert .
\end{equation*}
Integrating over $0\le t,t^\prime\le T$ and applying definition (\ref{covariancepath}), one obtains
\begin{equation}\label{encadrement1}
\vert (f, \hat{T}_{x^\prime(\cdot)}f)-(f, \hat{T}_{x(\cdot)}f)\vert
\le \frac{\varepsilon}{T}\left(\int_0^T\vert f(t)\vert\, dt\right)^2\le\varepsilon\| f\|_{L^2([0,T];\mathbb{R})}=\varepsilon .
\end{equation}
Now, let $f$ be a normalized eigenfunction of $\hat{T}_{x^\prime(\cdot)}$ associated with its largest eigenvalue $\mu_1[x^\prime(\cdot)]$. Then, $(f, \hat{T}_{x^\prime(\cdot)}f)=\mu_1[x^\prime(\cdot)]$, and  by the min--max principle \cite{CH1989,RS1978}, $(f, \hat{T}_{x(\cdot)}f)\le\mu_1[x(\cdot)]$. Substituting into equation (\ref{encadrement1}) gives $\mu_1[x^\prime(\cdot)]-\mu_1[x(\cdot)]\le\varepsilon$. Repeating the same argument with the roles of $x(\cdot)$ and $x^\prime(\cdot)$ exchanged yields $\mu_1[x(\cdot)]-\mu_1[x^\prime(\cdot)]\le\varepsilon$. Hence,
\begin{equation*}
\vert\mu_1[x(\cdot)]-\mu_1[x^\prime(\cdot)]\vert\le\varepsilon ,
\end{equation*}
which completes the proof of lemma \ref{mucontinuity}.

\bigskip
We now proceed to the proof of proposition \ref{criticalcoupling}. Since $\hat{T}_{x(\cdot)}$ is a compact and self-adjoint operator in $L^2([0,T];\mathbb{R})$, there exists an orthonormal basis $\lbrace \varphi_n\rbrace_{n\ge 1}$ such that $\hat{T}_{x(\cdot)}\varphi_n=\mu_n[x(\cdot)]\varphi_n$. Consider the sequence of random variables $Y_n=(S,\varphi_n)$. As linear functionals of the Gaussian field $S$, the $Y_n$'s are independent, zero-mean Gaussian random variables with $\langle Y_n^2\rangle_S=(\varphi_n,\hat{T}_{x(\cdot)}\varphi_n)=\mu_n[x(\cdot)]$. Decomposing $S$ on the orthonormal basis $\lbrace \varphi_n\rbrace_{n\ge 1}$ yields $\| S\|_{L^2([0,T],\mathbb{R})}^2=\sum_{n=1}^{+\infty}Y_n^2$. Injecting this decomposition into (\ref{pathampli}) and using the identities $<\exp(gY_n^2)>_S=1/\sqrt{1-2g\mu_n[x(\cdot)]}$, for $2g\mu_n[x(\cdot)]<1$, and $<\exp(gY_n^2)>_S=+\infty$, for $2g\mu_n[x(\cdot)]\ge 1$, one finds that, for any given $x(\cdot)\in B(0,T)$,
\begin{equation}\label{pathampliestim}
\mathcal{A}[x(\cdot); g] =\left\lbrace
\begin{array}{cr}
\prod_{n=1}^{+\infty}\frac{1}{\sqrt{1-2g\mu_n[x(\cdot)]}}\le\exp\left(\frac{gT}{1-2g\mu_1[x(\cdot)]}\right)
,&\forall g<1/2\mu_1[x(\cdot)], \\
+\infty ,&\forall g\ge 1/2\mu_1[x(\cdot)].
\end{array}\right.
\end{equation}
The inequality in (\ref{pathampliestim}) follows from $-\ln(1-x)\le x/(1-x)$ and $\sum_{n=1}^{+\infty}\mu_n[x(\cdot)]=C(0,0)\, T$, with normalization $C(0,0)=1$.

Write $\mu_{\rm max}=\sup_{x(\cdot)\in B(0,T)}\mu_1[x(\cdot)]$. For every $g<1/2\mu_{\rm max}$, it follows from (\ref{averageampli}) and (\ref{pathampliestim}) that $\langle\mathcal{E}(0,g)\rangle_S \le \exp[gT/(1-2g\mu_{\rm max})] <+\infty$ and $g_c\ge 1/2\mu_{\rm max}$.

By lemma \ref{mucontinuity}, for every $\varepsilon >0$, there exists at least one path $y(\cdot)\in B(0,T)$ such that $\mu_1[y(\cdot)]\ge \mu_{\rm max}-\varepsilon/2$. Choose such a path. By lemma \ref{mucontinuity}, there exists $\alpha >0$ such that for all $x(\cdot)\in B(0,T)$ with $\| x(\cdot)-y(\cdot)\|_\infty \le\alpha$, we have $\vert\mu_1[x(\cdot)]-\mu_1[y(\cdot)]\vert\le\varepsilon/2$. Define the set $B_{y(\cdot),\alpha}(0,T)=\lbrace x(\cdot)\in B(0,T) : \| x(\cdot)-y(\cdot)\|_\infty \le\alpha \rbrace$. Then, for all $x(\cdot)\in B_{y(\cdot),\alpha}(0,T)$, $\mu_1[x(\cdot)] \ge\mu_1[y(\cdot)]-\varepsilon/2 \ge\mu_{\rm max}-\varepsilon$, and for every $g\ge 1/2(\mu_{\rm max}-\varepsilon)$, it follows from (\ref{pathampliestim}) that $\mathcal{A}[x(\cdot); g]= +\infty$. Finally, since $B_{y(\cdot),\alpha}(0,T)$ has a strictly positive Wiener measure, one finds that $\langle\mathcal{E}(0,g)\rangle_S= +\infty$ for all $g\ge 1/2(\mu_{\rm max}-\varepsilon)$, and $g_c <1/2(\mu_{\rm max}-\varepsilon)$.

Combining the results, one obtains
\begin{equation}\label{gcinterval}
\frac{1}{2\mu_{\rm max}}\le g_c <\frac{1}{2(\mu_{\rm max}-\varepsilon)},
\end{equation}
and taking $\varepsilon>0$ arbitrarily small completes the proof of proposition \ref{criticalcoupling}.
%
%
\section{Proof of lemma \ref{neededbound}}\label{newappB}
For every realization of $S$, write
\begin{equation}\label{defofM}
M\equiv M(S)=\sup_{x(\cdot)\in B(0,T)}\, \sup_{t\in [0,T]} S(x(t),t)^2.
\end{equation}
The estimate (\ref{neededboundeq}) follows readily from equation (\ref{FKformula}) in which $S(x(\tau),\tau)^2$ is bounded by $M$. Write $R= \|x(\cdot)\|_\infty$, where $x(\cdot)$ is a Brownian path, and $p_R(r)=d\textrm{Prob.}(R<r)/dr$. One has
\begin{equation}\label{Mproba}
\textrm{Prob.}(M>m)=\int_0^{+\infty}\, \textrm{Prob.}(M>m\, \vert\, r)\, p_R(r)\, dr,
\end{equation}
where $\textrm{Prob.}(\cdot\, \vert\, r)$ is shorthand for the conditional probability given $r\le R <r+dr$. Define $D_{r,T}=[-r,r]\times [0,T]$. Since, for any point $(x,t)\in D_{r,T}$, there is at least one path in $B(0,T)$ with $\|x(\cdot)\|_\infty =r$ that passes through $(x,t)$, $M$ on the right-hand side of (\ref{Mproba}) is equal to
\begin{equation}\label{defofMrT}
M_{r,T} =\sup_{(x,t)\in D_{r,T}} S(x,t)^2,
\end{equation}
and one is led to determine the large $m$ behavior of
\begin{equation}\label{Mprobabis}
\textrm{Prob.}(M>m)=\int_0^{+\infty}\, \textrm{Prob.}(M_{r,T} >m)\, p_R(r)\, dr.
\end{equation}
The results of \cite{Adler1981,Delmas1998} on the statistics of maxima of Gaussian fields apply to the class of $S$ defined by the conditions \ref{condition1} and \ref{condition2}. For a finite two-dimensional domain, like $D_{r,T}$ with fixed $r>0$, one finds that the distribution of $M_{r,T}$ is absolutely continuous with upper tail
\begin{equation}\label{MrTproba}
\textrm{Prob.}(M_{r,T} >m)\sim\frac{\sqrt{2\, \vert\Lambda\vert}}{\pi^{3/2}}\, rT\, m^{1/2}\, \textrm{e}^{-m/2}
\qquad (m\to +\infty),
\end{equation}
where $\vert\Lambda\vert = \textrm{Var}(\partial_x S(0,0))\, \textrm{Var}(\partial_t S(0,0)) - \textrm{Cov}(\partial_x S(0,0),\partial_t S(0,0))^2$. Substituting (\ref{MrTproba}) into equation (\ref{Mprobabis}) gives
\begin{equation}\label{Mprobabisasym}
\textrm{Prob.}(M>m)\sim \frac{T\, \sqrt{2\vert\Lambda\vert}}{\pi^{3/2}}\, \left(\int_0^{+\infty} r\, p_R(r)\, dr\right)
\, m^{1/2}\textrm{e}^{-m/2}\qquad (m\to +\infty).
\end{equation}
By definition, $p_R(r)$ is the $r$-derivative of the survival probability at time $T$ for a Brownian walker constrained by $ \|x(\cdot)\|_\infty <r$, which is obtained by solving the heat equation with absorbing boundary conditions at $x=\pm r$. The calculation presents no particular difficulties and we leave it as an exercise for the reader. One finds
\begin{equation}\label{pdfofr}
p_R(r)=\frac{\pi T}{r^3}\, \sum_{k=0}^{+\infty}(-1)^k\, (2k+1)\, \exp\left[-\frac{(2k+1)^2\pi^2 T}{8 r^2}\right] .
\end{equation}
The super-exponential decay of $p_R(r)$ as $r\to 0$ ensures that $r\, p_R(r)$ in (\ref{Mprobabisasym}) is integrable near $r=0$. To verify integrability as $r\to +\infty$, we need to determine the large $r$ behavior of $p_R(r)$. Using the Poisson summation formula \cite{SW1971,Cordoba1988}, one gets (see appendix \ref{app1})
\begin{equation}\label{pdfofrpoisson}
p_R(r)=2\sqrt{\frac{2}{\pi T}}\, \sum_{m=-\infty}^{+\infty} (-1)^{m+1}\left( m-\frac{1}{2}\right)
\, \exp\left[-\left( m-\frac{1}{2}\right)^2 \frac{2r^2}{T}\right] .
\end{equation}
In the limit $r\to +\infty$, the sum in (\ref{pdfofrpoisson}) is dominated by the two terms $m=0$ and $m=1$, which yields
\begin{equation}\label{largerpofr}
p_R(r)\sim 2\sqrt{\frac{2}{\pi T}}\, \exp\left( -\frac{r^2}{2T}\right) \qquad (r\to +\infty),
\end{equation}
so that $r\, p_R(r)\sim r\, \exp(-r^2/2T)$ is integrable as $r\to +\infty$, and therefore the integral on the right-hand side of (\ref{Mprobabisasym}) exists.

Finally, substituting (\ref{Mprobabisasym}) into $p_M(m)=-d\textrm{Prob.}(M>m)/dm$ leads to equation (\ref{uppertailpdf}) with
\begin{equation}\label{constantA}
A=\frac{T}{\pi^{3/2}}\, \sqrt{\frac{\vert\Lambda\vert}{2}}\, \left(\int_0^{+\infty} r\, p_R(r)\, dr\right),
\end{equation}
which completes the proof of lemma \ref{neededbound}. Note that lemma \ref{neededbound} implies $g_c\ge 1/2T$, i.e. $\mu_{\rm max}\le~T$. This is consistent with the estimate $\mu_1[x(\cdot)]\le\sum_{n=1}^{+\infty}\mu_n[x(\cdot)]=C(0,0)\, T$ valid for all $x(\cdot)$ in $B(0,T)$, and hence also for $\mu_{\rm max}$, together with $C(0,0)=1$.
%
%
\section{Application of the Poisson summation formula to $\bm{p_R(r)}$}\label{app1}
For a given $a>0$, define the function of $s\in\mathbb{R}$
\begin{equation}\label{eqapp1}
h(s,a)={\rm e}^{i\pi s} (2s+1)\, {\rm e}^{-a(2s+1)^2}.
\end{equation}
Its Fourier transform
\begin{equation}\label{eqapp2a}
\hat{h}(q,a)=\int_{-\infty}^{+\infty} h(s)\, {\rm e}^{-2i\pi qs} ds ,
\end{equation}
is given by
\begin{equation}\label{eqapp2b}
\hat{h}(q,a)
=\frac{1}{4}\, \left(\frac{\pi}{a}\right)^{3/2} {\rm e}^{i\pi (q+1)} \left( q-\frac{1}{2}\right)
\, {\rm e}^{-\frac{\pi^2}{4a}\left( q-\frac{1}{2}\right)^2} .
\end{equation}
The Poisson summation formula \cite{SW1971,Cordoba1988} implies
\begin{equation}\label{eqapp3}
\sum_{n\in\mathbb{Z}} h(n,a)=\sum_{m\in\mathbb{Z}} \hat{h}(m,a).
\end{equation}
The expression (\ref{pdfofr}) for $p_R(r)$ can be rewritten in terms of $h(s,a)$ as
\begin{equation}\label{eqapp4}
p_R(r)=\frac{\pi T}{2r^3}\, \sum_{n=-\infty}^{+\infty} h\left( n,\frac{\pi^2 T}{8r^2}\right) ,
\end{equation}
where we have used the symmetry of the sum (the contribution from $n<0$ equals that from $n\ge 0$ after the change of variable $n\to -n-1$). Applying the summation formula (\ref{eqapp3}) on the right-hand side of (\ref{eqapp4}) then yields
\begin{equation}\label{eqapp5}
p_R(r)=\frac{\pi T}{2r^3}\, \sum_{m=-\infty}^{+\infty} \hat{h}\left( m,\frac{\pi^2 T}{8r^2}\right) ,
\end{equation}
which, in view of (\ref{eqapp2b}), coincides with equation (\ref{pdfofrpoisson}).
%
%

%
%
\end{document}